\documentstyle[12pt]{article}
\begin{document}

\author{V.B. Svetovoy\thanks{%
To whom correspondence should be addressed. E-mail: svetovoy@nordnet.ru} and
M.V. Lokhanin  \\ 
{\small Department of Physics, Yaroslavl State University,} \\ {\small %
Sovetskaya 14, Yaroslavl 150000, Russia}}
\title{Do the precise measurements of the Casimir force agree with the expectations?}
\date{}
\maketitle

\begin{abstract}
An upper limit on the Casimir force is found using the dielectric functions
of perfect crystalline materials which depend only on well defined material
constants. The force measured with the atomic force microscope is larger
than this limit at small separations between bodies and the discrepancy is
significant. The simplest modification of the experiment is proposed
allowing to make its results more reliable and answer the question if the
discrepancy has any relation with the existence of a new force.
\end{abstract}


The Casimir force \cite{Casimir} between closely spaced macroscopic bodies
is the effect of quantum electrodynamics (QED) and for that reason could be
predicted very accurately. In the rigorous Lifshitz theory \cite{Lif,LP} the
force is defined by the optical properties of used materials. Knowledge of
these properties is the weakest element in the theory restricting the
accuracy that can be achieved. Though the measurement of the Casimir force
is not the best way to test QED, such experiments are of great importance
because they are sensitive to the presence of new fundamental forces \cite
{Kuz} predicted in many modern theories (see, for example, \cite{Long} and
references therein). To distinguish a new force from the background, we
should be able to calculate the Casimir force with a precision better than
the experimental one. In the series of recent experiments this force has
been measured with the torsion pendulum (TP) \cite{Lam1} in the range of
distances $0.6-6\ \mu m$ and with the atomic force microscope (AFM) \cite
{MR,RLM} in the range $0.1-0.9\ \mu m$. The corresponding precisions were
5\% and 1\%, respectively.

The force per unit area between parallel plates arising as a result of
electromagnetic fluctuations at nonzero temperature $T$ is generalized by
the Lifshitz theory \cite{LP}, where the plate material is taken into
account by its dielectric function at imaginary frequencies $\varepsilon
\left( i\zeta \right) $:

\begin{equation}
\label{Liff}F^{pl}(a)=\frac{kT}{\pi c^3}{\sum\limits_{n=0}^\infty {}}%
^{\prime }\zeta _n^3\int\limits_1^\infty dpp^2\left\{ \left[ G_1^2e^{2p\zeta
_na/c}-1\right] ^{-1}+\left[ G_2^2e^{2p\zeta _na/c}-1\right] ^{-1}\right\} . 
\end{equation}

\noindent Here ''prime'' means that $n=0$ term is taken with the coefficient 
$1/2$, $a$ is the distance between bodies and

$$
G_1=\frac{p+s}{p-s},\quad G_2=\frac{\varepsilon \left( i\zeta _n\right) p+s}{%
\varepsilon \left( i\zeta _n\right) p-s},\quad 
$$

\begin{equation}
\label{defin1}s=\sqrt{\varepsilon \left( i\zeta _n\right) -1+p^2},\quad
\zeta _n=2\pi nkT/\hbar . 
\end{equation}

\noindent The Casimir result $F_c^{pl}\left( a\right) =\pi ^2\hbar c/240a^4$ 
\cite{Casimir} is reproduced from (\ref{Liff}) in the limit $\varepsilon
\rightarrow \infty $ and $T\rightarrow 0$. The function $\varepsilon \left(
i\zeta _n\right) $ cannot be measured directly but can be expressed via
imaginary part of the dielectric function on the real axis with the help of
the dispersion relation

\begin{equation}
\label{disp}\varepsilon \left( i\zeta \right) -1=\frac 2\pi
\int\limits_0^\infty d\omega \frac{\omega Im\varepsilon \left( \omega
\right) }{\omega ^2+\zeta ^2}. 
\end{equation}

\noindent Information on $Im\varepsilon \left( \omega \right) $ can be
extracted from the data on reflectivity and absorptivity of electromagnetic
waves for a given material.

In the experiments the force is measured between metalized disc and sphere
because for two plates it is difficult to keep them parallel. For this
configuration (\ref{Liff}) has to be modified with the help of the proximity
force theorem (PFT) \cite{PFT} which is true for $R\gg a$, where $R$ is the
radius of curvature of the spherical surface. Applying PFT to (\ref{Liff})
one can find the force between sphere and plate as $2\pi R\int F^{pl}\left(
a\right) da$. The integration gives

\begin{equation}
\label{shpl}F(a)=-\frac{kTR}{c^2}{\sum\limits_{n=0}^\infty {}}^{\prime
}\zeta _n^2\int\limits_1^\infty dpp\ln \left[ \left( G_1^{-2}e^{-2p\zeta
_na/c}-1\right) \left( G_2^{-2}e^{-2p\zeta _na/c}-1\right) \right] . 
\end{equation}

\noindent This expression differs from those used in \cite{RLM} and \cite
{Lam2} in two respects. First, in the cited papers the integration connected
with the PFT was not done analytically that complicated numerical analysis.
Second, the zero temperature limit has been taken when one can change the
sum over $n$ in (\ref{shpl}) by the integral over $\zeta $. This limit was
also considered in \cite{LR}, though the PFT integral was evaluated
explicitly. It seems a reasonable approximation at small separations because
the temperature correction is proportional to $(kTa/\hbar c)^3$ \cite{Temp}
and is small. However, one should remember that this correction has been
found in the limit of ideal conductor $\varepsilon \rightarrow \infty $. For
a real conductor it can behave as $kTa/\hbar c$ and be important. We have
computed the force according to (\ref{shpl}) and with the integral instead
of the sum and found that the difference at the smallest distances tested in
the AFM experiments exceeds $4\ pN$ in contrast with the conservative
estimate for the experimental errors $2\ pN$ \cite{RLM}.

In the AFM experiments an additional $Au_{0.6}Pd_{0.4}$ layer of $20\ nm$ 
\cite{MR} or $8\ nm$ \cite{RLM} thick was on the top of $Al$ metallization
of the bodies to prevent aluminum oxidation. It has to be included into
consideration. This layer is transparent for the electromagnetic waves with
high frequencies $\sim c/a$ since the absorption is proportional to $%
Im\varepsilon \left( \omega \right) $ which is small for $\omega \sim c/a$
and for this reason the layer was ignored in \cite{MR,RLM}. However, the
force depends on $\varepsilon (i\zeta )$ for which the low frequencies
dominate in (\ref{disp}) because of large $Im\varepsilon \left( \omega
\right) $ and that is why we cannot neglect the $Au/Pd$ layer. To take it
into account, one has to generalize expression for the force (\ref{Liff}) to
the case of layered bodies. Suppose that the top layer has the thickness $h$
and its dielectric function is $\varepsilon _1$. The bottom layer is thick
enough to be considered as infinite and let its dielectric function be $%
\varepsilon _2$. The method described in \cite{LP} for deriving Eq.(\ref
{Liff}) can be easily generalized for layered plates. We have to add only
the matching conditions for the Green functions on the layers interface. The
result will look exactly as (\ref{Liff}) but with more complex $G_{1,2}$:

$$
G_1=\frac{\left( s_1+s_2\right) \left( p+s_1\right) e^{\zeta
_ns_1h/c}+\left( s_1-s_2\right) \left( p-s_1\right) e^{-\zeta _ns_1h/c}}{%
\left( s_1+s_2\right) \left( p-s_1\right) e^{\zeta _ns_1h/c}+\left(
s_1-s_2\right) \left( p+s_1\right) e^{-\zeta _ns_1h/c}}, 
$$

\begin{equation}
\label{defin2}G_2=-\frac{\left( \varepsilon _2s_1+\varepsilon _1s_2\right)
\left( \varepsilon _1p+s_1\right) e^{\zeta _ns_1h/c}+\left( \varepsilon
_2s_1-\varepsilon _1s_2\right) \left( \varepsilon _1p-s_1\right) e^{-\zeta
_ns_1h/c}}{\left( \varepsilon _2s_1+\varepsilon _1s_2\right) \left(
\varepsilon _1p-s_1\right) e^{\zeta _ns_1h/c}+\left( \varepsilon
_2s_1-\varepsilon _1s_2\right) \left( \varepsilon _1p+s_1\right) e^{-\zeta
_ns_1h/c}}, 
\end{equation}

\noindent where $s_{1,2}$ are defined similar to $s$ in (\ref{defin1}). The
force between plate and sphere is given by (\ref{shpl}) with the above $%
G_{1,2}$. Qualitatively the effect of the top layer will be negligible if $%
h\omega _{1p}/c\ll 1$ where $\omega _{1p}$ is the plasma frequency of this
layer. For typical plasma frequencies $\sim 10^{16}\ s^{-1}$ it is
definitely not the case even for $h=8\ nm$. The force between layered bodies
was found also in \cite{LR} with a little bit different technic but it was
not used there for actual calculations.

Now we are able to evaluate the Casimir force in real geometry of the
experiments if there is information on the dielectric functions of the used
materials: $Au$, $Al$, and $Au_{0.6}Pd_{0.4}$ alloy. Strictly speaking, one
has to measure these functions in wide range of wavelengths on the samples
which are used for the force measurement. It was not done in all experiments
and to draw any conclusion from them we have to make some assumptions on $%
\varepsilon \left( \omega \right) $. At low frequencies $Au$ and $Al$ are
well described by the Drude theory, where 
\begin{equation}
\label{Drudr}\varepsilon \left( \omega \right) =1-\frac{\omega _p^2}{\omega
\left( \omega +i\omega _\tau \right) }. 
\end{equation}

\noindent Here $\omega _p$ is the free electron plasma frequency and $\omega
_\tau $ is the Drude damping frequency. A simple test for validity of (\ref
{Drudr}) is the behavior of the material resistivity \cite{Pers} that is
defined as

\begin{equation}
\label{res}\rho \left( \omega \right) =Im\frac 1{\varepsilon _0\left(
1-\varepsilon \left( \omega \right) \right) \omega }=\frac{\omega _\tau }{%
\varepsilon _0\omega _p^2}, 
\end{equation}

\noindent where $\varepsilon _0$ is the free space permittivity. The
resistivity is frequency independent within the Drude approximation. For
crystalline samples of $Au$ and $Al$ (the entries 2 in Table 1) the
frequency behavior of the resistivity and $Im\varepsilon \left( \omega
\right) $ are shown in Fig.1. The data on the dielectric functions were
taken from \cite{Zol}, where the data from many original works are
collected. Palladium definitely cannot be described by (\ref{Drudr}) at any
frequency. However, it is known experimentally that amorphous metallic
alloys such as $Au/Pd$ can be described by the Drude approximation \cite
{Pers}. The physical explanation for this is associated with large Drude
damping of the compounds like $Au_{0.6}Pd_{0.4}$. Eq.(\ref{res}) allows to
use well defined static resistivity $\rho \left( 0\right) =\rho _0$ instead
of the damping frequency $\omega _\tau $.

Of course, at higher frequencies when interband transitions are reached the
Drude approximation fails. Nevertheless, it is very useful since low
frequencies dominate in the dispersion relation. Extrapolation of (\ref
{Drudr}) to high frequencies gives

\begin{equation}
\label{dfimag}\varepsilon \left( i\zeta \right) =1+\frac{\omega _p^2}{\zeta
\left( \zeta +\omega _\tau \right) }. 
\end{equation}

\noindent The relative error inserted in (\ref{dfimag}) due to extrapolation
can be estimated as $\omega _\tau /\omega _0$, where $\omega _0$ is the
frequency of the first resonance for a given metal. The error can be as
large as 10\% but it does not influence significantly in the force. If we
will use (\ref{dfimag}) for the force computation and change $\omega _p$ by
5\% ( 10\% correction to $\varepsilon \left( i\zeta \right) $ at all
frequencies), then the force is changed less than 2\%. Moreover, since the
interband transitions give a correction to (\ref{dfimag}) which is frequency
dependent, it reduces the correction to the force further below the
experimental uncertainties. The possibility to neglect the interband
transitions in $Al$ for the force evaluation was noted in \cite{RLM}. It
agrees with our estimate and direct computation using the handbook data for $%
Im\varepsilon \left( \omega \right) $. Therefore, in all cases of interest
we can use Eq.(\ref{dfimag}) to describe the dielectric function of a
material on the imaginary axis. Since the integral in (\ref{disp}) is
saturated in the low frequency region, we should extract the parameters $%
\omega _p$ and $\omega _\tau $ from the data for real and imaginary parts of 
$\varepsilon \left( \omega \right) $ by fitting them in infrared region with
(\ref{Drudr}).

It is important that optical properties of evaporated (spattered) films can
be quite different from those of bulk material and depend on the
technological details. It is known, for example, that the film density is
typically 0.7 of that of the bulk material if it was not annealed. For the
resistivity of spattered and evaporated $Au$ films the value $\rho _0=8.2\
\mu \Omega \cdot cm$ has been reported \cite{Audat} in contrast with the
bulk resistivity $2.25\ \mu \Omega \cdot cm$. All these make impossible to
use the handbook data for reliable calculation of the Casimir force. This
conclusion is illustrated by Table \ref{Tab1}, where the parameters for $Al$
and $Au$ found by fitting the data from \cite{Zol} are presented.

\begin{table} \centering
  \begin{tabular}{|c|c|c|c|} \hline
     {\bf Al}  &  $\omega_p \cdot 10^{-16} $  &  $\omega_{\tau} %
     \cdot 10^{-13} $  &  $\rho_0 \ \mu \Omega \cdot cm$ \\ \hline
     $1^*$       & $1.54\pm 0.01$ & $15.5\pm 0.6$  & 7.39 \\
     2       & $2.235\pm 0.001$ & $12.49\pm 0.01$ & 2.83 \\
     3       & $2.43\pm 0.05$ & $14.4\pm 0.7$ & 2.76 \\ 
     $4^*$       & $1.63\pm 0.03$ & $18.2\pm 0.7$ & 7.74 \\ \hline
      {\bf Au}  &  $\omega_p \cdot 10^{-16} $  &  $\omega_{\tau} %
      \cdot 10^{-13} $  &  $\rho_0 \ \mu \Omega \cdot cm$ \\ \hline
     1       & $1.280\pm 0.001$ & $3.29\pm 0.05$  & 2.27 \\
     2       & $1.372\pm 0.001$ & $4.060\pm 0.002$ & 2.44\\
     3       & $1.34\pm 0.02$ & $7.08\pm 0.18$ & 4.46\\ 
     4       & $1.051\pm 0.001$ & $6.24\pm 0.21$ & 6.40 \\ \hline
  \end{tabular}
  \caption{Drude parameters with statistical errors for $Al$ and $Au$ 
              found by fitting $\epsilon(\omega)$ in the infrared range 
              ($\lambda > 2 \ \mu m$). The stars in the first column mark 
              the data for film samples.\label{Tab1}}
\end{table}

Though we cannot use handbook data to evaluate the force, one can confine it
for a given experiment. This statement is based on the observation that
because of better reflectivity the force (\ref{shpl}) increases every time
when $\omega _p$ increases or $\rho _0$ decreases. For us it is important
that any technological procedures will reduce $\omega _p$ and increase the
resistivity $\rho _0$. The perfect crystalline material will have the
largest plasma frequency and the smallest resistivity and these parameters
are well defined. The plasma frequency $\omega _p$ is defined by the
concentration of free electrons in the metal $n$ and their effective mass $%
m_e^{*}$

\begin{equation}
\label{omp}\omega _p=\sqrt{\frac{e^2n}{m_e^{*}\varepsilon _0}}. 
\end{equation}

\noindent Gold is a good conductor and $m_e^{*}$ is quite close to the mass
of electron. We will find the upper limit on the electron concentration if
suppose that every $Au$ atom produce a free electron. Then for $Au$ plasma
frequency one finds $\omega _p^{Au}=1.37\cdot 10^{16}\ s^{-1}$. The static
resistivity can be used to get the damping frequency $\omega _\tau $ with
the help of (\ref{res}) at a given $\omega _p$. For crystalline gold it is $%
\rho _0^{Au}=2.25\ \mu \Omega \cdot cm$. One can compare these parameters
with that given in Table 1 to make sure that they correspond to the limit
values. In the TP experiment \cite{Lam1} the bodies were covered with $Au$
of thickness $0.5\ \mu m$ that is thick enough to be considered as infinite.
Substituting the $Au$ parameters into (\ref{dfimag}) and calculating the
force according to (\ref{shpl}) one finds the upper limit on the Casimir
force in the TP experiment. The residual force $F^{exp }\left( a_i\right)
-F^{lim }(a_i)$ is shown in Fig.2a, where $F^{exp }\left( a_i\right) $ are
the experimental points at separations $a_i$. The prediction obviously does
not contradict to the experiment but dealing with the upper limit we cannot
conclude that there is an agreement, either.

For the AFM experiments \cite{MR,RLM} the upper limit on the Casimir force
is more restrictive. The plasma frequency for $Al$ can be restricted using (%
\ref{omp}) if one supposes that every atom produces 3 free electrons. It
gives $\omega _p^{Al}=2.40\cdot 10^{16}\ s^{-1}$ that coincide with the
largest value in Table \ref{Tab1}. The resistivity of perfect crystal is $%
\rho _0^{Al}=2.65\ \mu \Omega \cdot cm$. Since we successfully predicted the
plasma frequencies for the best samples of $Au$ and $Al$, the same way one
can use to estimate $\omega _p$ for $Au/Pd$. If each $Au$ atom gives one and 
$Pd$ atom gives not more than two free electrons, then $\omega
_p^{Au/Pd}=1.69\cdot 10^{16}\ s^{-1}$. This alloy is used in
microelectronics and resistivity of the bulk material is known to be $\rho
_0^{Au/Pd}\approx 30\ \mu \Omega \cdot cm$ \cite{Kriv} in accordance with
the statement that alloys have large Drude damping. These data allow to find
the upper limit on the force using (\ref{shpl}) with the functions $G_{1,2}$
defined in (\ref{defin2}). Real surface of the bodies is always distorted.
The distortion statistics were analysed with atomic force microscope \cite
{RLM,KRMM}. The force has to be averaged over the distorted surfaces and we
use for this the procedure developed in \cite{KRMM}. This procedure seems
quite reliable. Moreover, the important progress in controlled metal
evaporation \cite{RLM} allowed to reduce the surface roughness to the level
when the correction to the force becomes practically unimportant for the
experiment \cite{RLM}.

It was indicated \cite{MR} that the thickness of $Au/Pd$ layer is less than $%
20\ nm$, that is why for calculations the conservative value $h=15\ nm$ was
chosen. The top layer changes the force on $13\ pN$ at the smallest
separation. Variation of $\omega _p^{Al}$ on 10\% gives only $1\ pN$ change
in the force because of screening effect of the top layer. The same
variation in $\omega _p^{Au/Pd}$ changes the force on $2\ pN$. The
resistivity variation of the $Au/Pd$ layer on 30\% gives $1\ pN$ effect. At
larger separations all the effects become smaller. All this means that the
limit is stable in respect to variation of the parameters. It is clear also
that the top layer definitely cannot be ignored in the force evaluation. The
residual force $F^{exp }\left( a\right) -F^{lim }(a)$ with the experimental
points from \cite{MR} is shown in Fig.2b by triangles.

In \cite{RLM} the assumption of absolute transparency of the $Au/Pd$ layer
was not only used for theoretical interpretation of the result but also in
the procedure of the force extraction from the raw data. For this reason we
cannot use the points for the force directly. Fortunately, it is easy to
restore the right data by shifting all the points to larger separations on $%
2h=16\ nm$. The result for the residual force $F^{exp }\left( a\right)
-F^{lim }(a)$ with shifted experimental points from \cite{RLM} is presented
in Fig.2b by open squares. This figure clearly indicates the presence of
some unexplained attractive force which is decreasing rapidly when the
distance between bodies increased. One can speculate that the observed
discrepancy is explained by a new Yukawa force mediated by a light scalar
boson but we will not discuss now the restrictions on the Yukawa parameters
that will be given elsewhere.

To make the experiment absolutely clear, it is preferable to use $Au$
instead of $Al$ metallization because its non-reactive surface has strong
advantage over $Al$. It excludes also additional uncertainties connected
with $Au/Pd$ layer. One can use as well silver or copper but they are not as
inert as gold. In practice it is difficult to measure the dielectric
function at the wavelengths larger than $30\ \mu m$ but this range gives an
important contribution to the dispersion relation. That is why the material
behavior in this range has to be predictable. One can say definitely that
the materials of platinum group cannot be used since they are not described
by (\ref{Drudr}) at low frequencies. An additional advantage of $Au$
metallization is higher density of the bodies coating. In this case the
hypotetic Yukawa force will be larger roughly in $\left( \rho _{Au}/\rho
_{Al}\right) ^2\approx 50$ times. If the observed discrepancy has relation
with the Yukawa interaction, the AFM experiment with $Au$ metallization of
the bodies will definitely reveal this new force even without detailed
knowledge of optical properties of the metallization.

In conclusion, we have found the upper limit on the Casimir force that is
realized for perfect crystalline coating of the bodies for which electrical
and optical properties are well defined. This limit is smaller than the
observed force in the AFM experiments and the difference far exceeds
experimental errors and theoretical uncertainties for small separations
between bodies. $Au$ metallization of the bodies in the AFM experiment will
allow to reveal origin of the discrepancy.

\newpage

\centerline{\large {\bf Figure captions}}

\vspace{0.5cm}

\noindent {\bf Figure 1}. Validity of the Drude approximation for $Al$
(triangles) and $Au$ (circles) in the infrared range. The resistivity does
not depend on frequency (left axis). Solid lines (right axis) demonstrate
that $Im\varepsilon \left( \omega \right) $ depends on $\omega $ according
to (\ref{Drudr}) with the parameters given in Table \ref{Tab1} (entries 2).

\vspace{0.5cm}

\noindent {\bf Figure 2}. The residual force $F^{exp }\left( a_i\right)
-F^{lim }(a_i)$ for different experiments: (a) TP experiment \cite{Lam1};
(b) AFM experiments with the data from \cite{MR} (triangles) and from \cite
{RLM} (open squares).


\begin{thebibliography}{99}
\bibitem{Casimir}  H.B.G. Casimir, Koninkl. Ned. Adak. Wetenschap. Proc. 
{\bf 51}, 793 (1948).

\bibitem{Lif}  E.M. Lifshitz, Sov. Phys. JETP {\bf 2}, 73 (1956).

\bibitem{LP}  E.M. Lifshitz and L.P. Pitaevskii, {\it Statistical Physics,
Part 2}, (Pergamon Press, Oxford, 1980).

\bibitem{Kuz}  V.A. Kuz'min, I.I. Tkachev, and M.E. Shaposhnikov, JETP
Letters (USA) {\bf 36}, 59 (1982).

\bibitem{Long}  J.C. Long, H.W. Chan, and J.C. Price, Nucl. Phys. {\bf B539}%
, 23 (1999), hep-ph/9805217.

\bibitem{Lam1}  S.K. Lamoreaux, Phys. Rev. Lett. {\bf 78}, 5 (1997); {\bf 81}%
, 5475 (1998).

\bibitem{MR}  U. Mohideen and A. Roy, Phys. Rev. Lett. {\bf 81}, 4549
(1998), physics/9805032.

\bibitem{RLM}  A. Roy, C.-Y. Lin, and U. Mohideen, Phys. Rev. D{\bf \ 60},
111101 (1999), quant-ph/9906062.

\bibitem{PFT}  J.Blocki, J. Randrup, W.J. Swiatecki, and C.F. Tsang, Ann.
Phys. (N.Y.) {\bf 105}, 427 (1977).

\bibitem{Temp}  L.S. Brown and G.J. Maclay, Phys. Rev. {\bf 184}, 1272
(1969).

\bibitem{KRMM}  G.L. Klimchitskaya, A. Roy, U. Mohideen, and V.M.
Mostepanenko, Phys. Rev. A{\bf \ 60}, 3487 (1999), quant-ph/9906033.

\bibitem{Lam2}  S.K. Lamoreaux, Phys. Rev. A{\bf \ 59} R3149 (1999).

\bibitem{LR}  A. Lambrecht and S. Reynaud, Eur. Phys. J. {\bf D}, to appear,
quant-ph/9907105.

\bibitem{Pers}  B.N.J. Persson, J.E. Demuth, Phys. Rev. B{\bf \ 31}, 1856
(1985).

\bibitem{Zol}  V.M. Zolotarev, V.N. Morozov, and E.V. Smirnov, {\it Optical
constants of natural and technical mediums }(Chimiya, Leningrad,1984).

\bibitem{Audat}  L. Lechevallier, G. Richon, J. Le Bas, and M. Bernole,
Vacuum {\bf 41}, 1218 (1990).

\bibitem{Kriv}  {\it Materials in mechanical engineering and automation}. 
{\it Handbook. }Edited by{\it \ } Yu.M. Pyatin, 2nd ed. (Mashinostroenie,
Moskva, 1982)
\end{thebibliography}
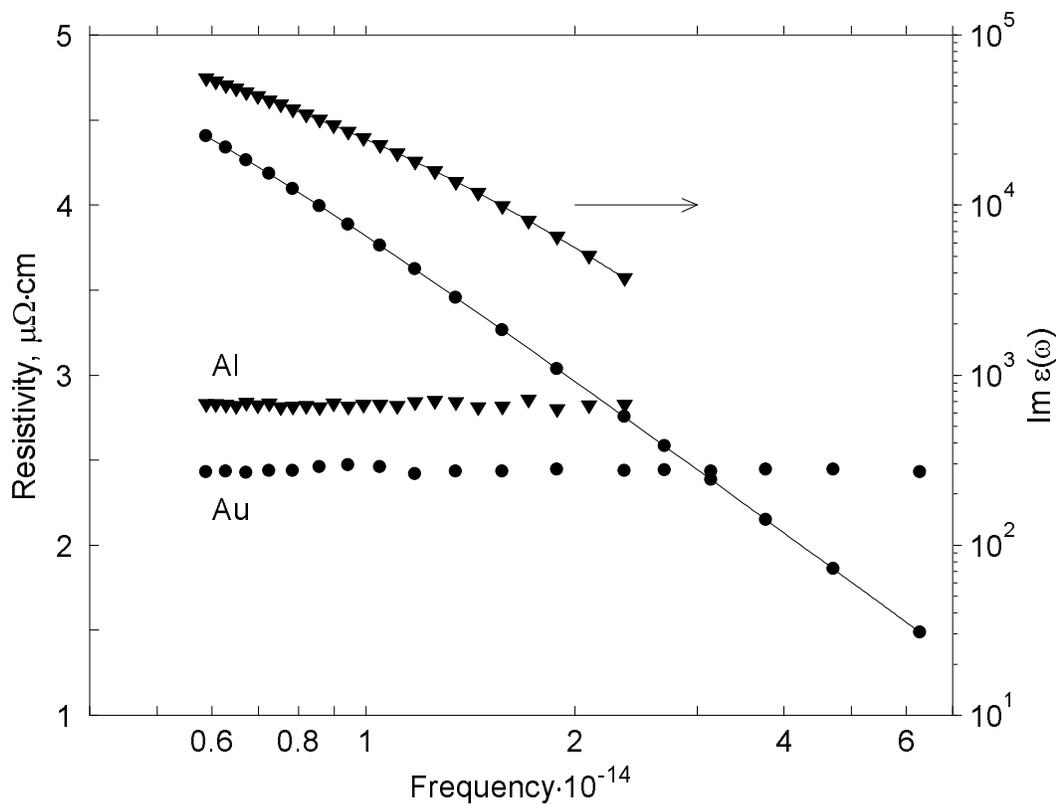
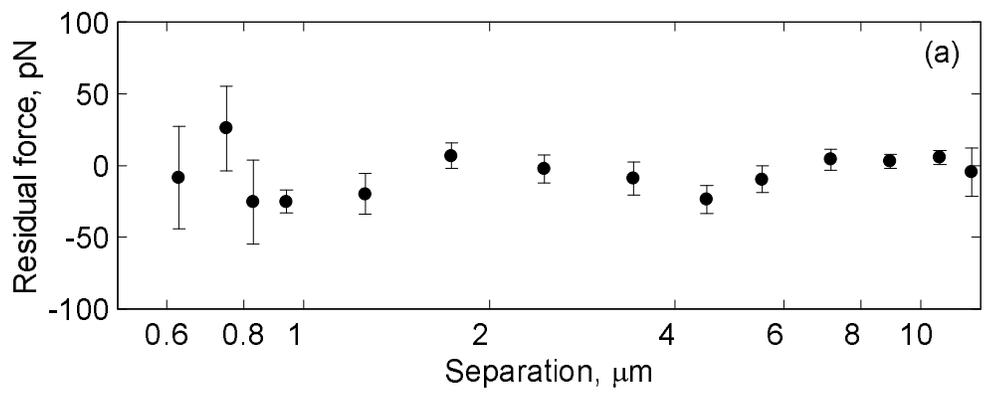
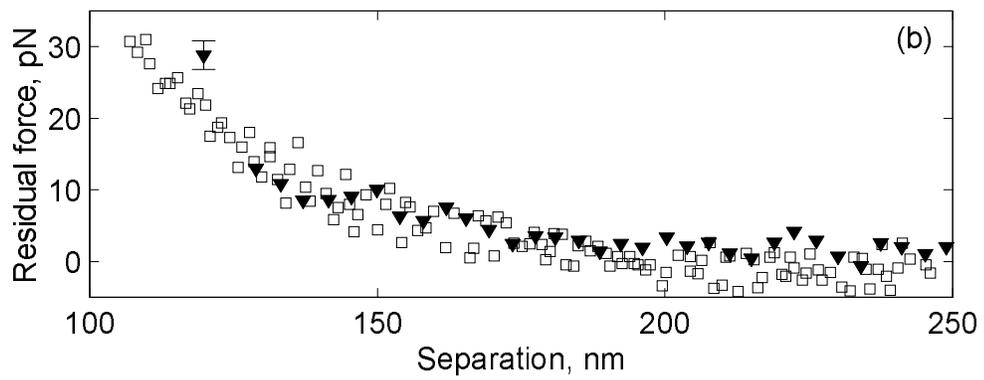
\end{document}